\documentclass[prd,aps,preprint,nofootinbib,superscriptaddress]{revtex4}
\pdfoutput=1
\usepackage{amsmath}
\usepackage{epsfig}
\usepackage{graphicx}
\usepackage{amssymb}
\usepackage{subfigure}

\def\beqn{\begin{eqnarray}} 
\def\eeqn{\end{eqnarray}} 
\def\be{\begin{equation}}
\def\ee{\end{equation}}
\def\nn{\nonumber}

\begin{document}

\title{Cosmic Ray-Dark Matter Scattering: a New Signature of (Asymmetric) Dark Matter in the Gamma Ray Sky}

\author{Stefano Profumo}
\email{profumo@scipp.ucsc.edu} \affiliation{Department of Physics, University of California, 1156 High St., Santa Cruz, CA 95064, USA}\affiliation{Santa Cruz Institute for Particle Physics, Santa Cruz, CA 95064, USA} 
\author{Lorenzo Ubaldi}
\email{ubaldi@physics.ucsc.edu}
\affiliation{Department of Physics, University of California, 1156 High St., Santa Cruz, CA 95064, USA}

\date{\today}

\begin{abstract}
\noindent We consider the process of scattering of Galactic cosmic-ray electrons and protons off of dark matter with the radiation of a final-state photon. This process provides a novel way to search for Galactic dark matter with gamma rays. We argue that for a generic weakly interacting massive particle, barring effects such as co-annihilation or a velocity-dependent cross section, the  gamma-ray emission from cosmic-ray scattering off of dark matter is typically smaller than that from dark matter pair-annihilation. However, if dark matter particles {\em cannot} pair-annihilate, as is the case for example in asymmetric dark matter scenarios, cosmic-ray scattering with final state photon emission provides a unique window to detect a signal from dark matter with gamma rays. We estimate the expected flux level and its spectral features for a generic supersymmetric setup, and we also discuss dipolar and luminous dark matter. We show that in some cases the gamma-ray emission might be large enough to be detectable with the Fermi Large Area Telescope.
\end{abstract}

\maketitle

\section{Introduction}

Indirect dark matter detection is the search for the annihilation or decay products of particle dark matter --- instead of {\em directly} detecting the dark matter, {\em indirect} detection aims at pinpointing the non-standard origin of a cosmic-ray or a gamma-ray signal based upon the messenger's spectral or angular features (see the relevant chapters of Ref.~\cite{2010pdmo.book.....B} for a collection of recent reviews). Examples of indirect detection with gamma rays are searches for a peculiar gamma-ray signal from nearby dwarf spheroidal galaxies (see e.g. \cite{Abdo:2010ex}), from nearby clusters (e.g. \cite{Jeltema:2008vu, Ackermann:2010rg, Dugger:2010ys}), from the center of the Galaxy (e.g. \cite{Jeltema:2008hf}), in the extragalactic gamma-ray background (e.g. \cite{Abdo:2010dk, Profumo:2009uf}) or in the diffuse background angular distribution (e.g. \cite{Ando:2005xg}); high-energy neutrinos from the center or of the Sun or of the Earth (see e.g.  \cite{Ellis:1987cy, Brown:2010cs}), the search for features in the spectrum of positrons or antiprotons (e.g. \cite{Profumo:2004ty}), and the search for antidetuerons (e.g. \cite{Mori:2001dv, Baer:2005tw}) are further examples of indirect dark matter detection channels.

This is an especially exciting point in time for the field of indirect dark matter detection. The Fermi Large Area Telescope (LAT) is well into its third, successful year of science operations. Crucial aspects of the gamma-ray sky for indirect dark matter searches, such as the issue of the galactic and extra-galactic diffuse background, are undergoing close scrutiny with significant progress well under-way \cite{Abdo:2009mr, Abdo:2010nz}. Research and development is also in progress for the next generation gamma-ray detector, a ground-based Cherenkov Telescope Array (CTA), whose significantly lower low-energy threshold will enormously benefit its use for dark matter searches. The IceCube neutrino telescope has recently been completed, with a thickly-instrumented portion (DeepCore) that makes it an even better dark matter detection instrument than originally designed \cite{Brown:2010cs}. Finally, the AMS-02 payload has recently been successfully deployed on-board the International Space Station, and science results will soon be delivered.

The weakly-interacting massive particle (WIMP) paradigm for the nature of dark matter is a highly compelling scenario, as it relies on generic weak-scale physics and on the well-known thermal freeze-out mechanism for the production of a relic with just the right density to match the inferred dark matter abundance in the universe. However, the so-called WIMP miracle does not provide an explanation to why the universal density of baryons $\Omega_{\rm B}$ (in units of the critical density, $\Omega_{\rm B}\equiv\rho_{\rm B}/\rho_{\rm crit}$) is tantalizingly close to the density of dark matter $\Omega_{\rm DM}$, so much as to pose the veritable ``coincidence problem'' of explaining why $\Omega_{\rm DM}/\Omega_{\rm B}\sim5$. A general class of models posits that the dark matter relic density is in fact set by an asymmetry between dark matter and anti-dark matter particles, in full similarity to the baryonic relic density (e.g. \cite{Gelmini:1986zz, Barr:1991qn, Kaplan:1991ah, Dodelson:1991iv}). If the two asymmetries are related to each other (for example if there exists a ``transfer mechanism'' between the dark matter and the baryonic sector), then generically $m_{\rm DM}\approx c\ (\Omega_{\rm DM}/\Omega_{\rm B})m_p\approx 5$ GeV, with $m_p$ the proton mass and $c$ a model-dependent number of order unity. These models have been dubbed ``asymmetric dark matter'' models in Ref.~\cite{Kaplan:2009ag}, and received a considerable amount of attention recently, in part because of direct detection results that might point to a light (with a mass of a few GeV) dark matter candidate.

While asymmetric dark matter models give rise to an interesting phenomenology (see e.g. \cite{Kaplan:2009ag, Shelton:2010ta}), including new contributions to proton decay, and, as mentioned, can even explain anomalies in direct dark matter experiments (e.g. \cite{Fitzpatrick:2010em}), by definition, asymmetric dark matter does not annihilate (or does so at a very suppressed rate, should there be a left-over symmetric component). One might thus naturally wonder whether indirect dark matter detection is or not entirely doomed in asymmetric dark matter scenarios (see however \cite{Cai:2009ia, Chang:2011xn, Graesser:2011wi, Feldstein:2010xe}).

In this paper we consider a process that was, in an entirely different astrophysical setting, suggested some time ago by Bloom and Wells \cite{Bloom:1997vm}. There, the Authors suggested that high-energy electrons in the jets of active galactic nuclei (AGN) could up-scatter dark matter particles, possibly radiating a photon in the final state. Estimating, with the techniques of effective field theory, the relevant cross section, and estimating the flux of electrons and the density of dark matter particles, the authors concluded that (1) the level of the signal was several orders of magnitude below the sensitivity of the LAT, and (2) no specific ``smoking-gun'' spectral features were present in the signal.

In Ref.~\cite{Gorchtein:2010xa} we recently studied the same process originally proposed in \cite{Bloom:1997vm}, improving on the previous analysis in a number of ways. First, we carried out the complete elementary process calculation, including resonance and threshold effects, for two particle dark matter models (supersymmetric neutralinos and the lightest Kaluza-Klein of universal extra dimensions) for both cosmic-ray electrons and protons in the AGN jet; Second, we re-evaluated the AGN jet components and the dark matter density in view of recent observational and theoretical-modeling results. We concluded that the process can be resonant, and that the resonance entails on the one hand a larger final-state photon flux, and on the other hand a distinctive spectral feature, corresponding to a certain, well-understood kinematic condition. As a result, we predicted that for some dark matter models the process of dark matter scattering off of AGN jet particles can be observable by the Fermi LAT.

Here, we observe that our own Galaxy is populated by cosmic rays (hadrons and heavy nuclei, electrons, positrons and neutrinos) that might interact with the Galactic dark matter --- a process that might be ``visible'' with a gamma-ray telescope such as Fermi LAT if it radiates a photon in the final state. Evidently, the elementary processes we considered in detail in Ref.~\cite{Gorchtein:2010xa} are precisely the same for the case of scattering off of Galactic cosmic rays (albeit a few technical differences will be of importance, as we point out later). A further possibility, that we leave for future work, is that dark matter-cosmic ray scattering lead to inelastic processes producing e.g. neutral pions, eventually resulting in gamma rays. Note that inelastic interactions between cosmic-ray protons and dark matter particles have been also recently considered in Ref.~\cite{Masip:2008mk} and \cite{Barcelo:2009uy} as a possible explanation to the origin of the observed ``knee'' in the spectrum of galactic cosmic rays, for dark matter models in the context of extra dimensions with a low fundamental scale of gravity. Limits on the interaction cross section of dark matter off of hadronic cosmic rays have also been considered in Ref.~\cite{Cyburt:2002uw} in the context of models of strongly interacting dark matter; the strongest constraints were found to arise from effects on the predicted abundances of light elements in big bang nucleosynthesis.

In this study we first consider, in Sec.~\ref{rough}, the process of dark matter scattering off of cosmic rays in the Galaxy for a generic WIMP model, and we compare the expected luminosity from final-state radiated photons with what expected for WIMP pair-annihilation; we also consider the same process for two additional dark matter models that received some attention recently: dipolar dark matter and ``luminous'' dark matter. We then present the core of our numerical results in Sec.~\ref{core}, and separately give details for the case of cosmic-ray electrons (\ref{ele}) and protons (\ref{pro}). We present a discussion and our conclusions in the final Sec.~\ref{sec:disc}.

\section{Dark Matter Pair Annihilation versus Cosmic Ray-Dark Matter Scattering}\label{rough}

In this Section we estimate, for generic WIMP models as well as for dipolar and luminous dark matter models, the expected luminosity for the process where a photon is emitted in the final state of a cosmic-ray scattering off of dark matter. 

We start by comparing, for the case of (symmetric) WIMP dark matter models where pair-annihilation can occur, the luminosity of the final-state photon emission from cosmic ray scattering off of dark matter to the well-known dark matter pair-annihilation rate. Both processes can be thought of as occurring on the same target particles: Galactic dark matter. We thus construct an order-of-magnitude estimate for the relevant rates, per unit target. These are defined here as follows:
\begin{itemize}
\item for dark matter annihilation, $$ q_{\chi\chi}\equiv\sigma_{\chi\chi}\  n_\chi\ v_{\rm rel},$$ with $\sigma_{\chi\chi}$ the pair annihilation cross section, $n_\chi$ the number density, and $v_{\rm rel}$ the relative dark matter particle velocity in the Galaxy
\item for cosmic ray-dark matter scattering, $$ q_{e\chi\to e\chi\gamma}\equiv\left(E\frac{{\rm d}\phi_e}{{\rm d}E}\Big|_{1\ \rm GeV}\right)\times\sigma_{e\chi\to e\chi\gamma},$$ with $\frac{{\rm d}\phi_e}{{\rm d}E}$ the locally measured differential cosmic ray electron flux, averaged over angles, and $\sigma_{e\chi\to e\chi\gamma}$ the relevant scattering cross section.
\end{itemize}
We estimate a weak-scale pair-annihilation cross section $$\sigma_{\chi\chi}\sim\frac{\alpha^2}{m^2_{\rm EW}}\sim ({\rm few})\times 10^{-36}\ {\rm cm}^2,$$ and we employ a local dark matter density $\rho_\chi\sim0.3\ {\rm GeV\ cm}^{-3}$ and a dark matter mass of 100 GeV to estimate $n_\chi\sim3\times10^{-3}\ {\rm cm}^{-3}$. Finally, we take $v_{\rm rel}\sim10^{-3}c$. With these numbers, we find $$q_{\chi\chi}\sim({\rm few})\times 10^{-31}\ \frac{1}{\rm s}.$$
For the case of cosmic ray-dark matter scattering we employ $$E\frac{{\rm d}\phi_e}{{\rm d}E}\Big|_{1\ \rm GeV}\sim4\times 10^{-1}\frac{1}{{\rm cm}^{2}\ {\rm s}},$$ as measured and reported by the LAT collaboration \cite{Ackermann:2010ij} (we consider cosmic-ray electron energies around 1 GeV given that we are interested in final-state gamma rays with energies in the Fermi LAT range, thus above 0.1 GeV, and because at larger energies the cosmic-ray spectrum declines very steeply; we thus expect the energy range that contributes the most to the signal we are interested in to be in the GeV range), and we estimate $$\sigma_{e\chi\to e\chi\gamma}\sim\frac{\alpha^3}{m^2_{\rm EW}}\sim ({\rm few})\times 10^{-38}\ {\rm cm}^2,$$ with possible resonances enhancing these numbers by 2-4 orders of magnitude, according to the findings of \cite{Gorchtein:2010xa} (see e.g. Fig.~3). As a result, $$q_{e\chi\to e\chi\gamma}\sim10^{-38}\ \frac{1}{\rm s} \to ({\rm few})\times 10^{-34}\ \frac{1}{\rm s}.$$
This calculation only employs the local dark matter density and cosmic-ray electron flux: both these quantities are larger the closer one gets to the center of the Galaxy. This order of magnitude estimate, however, implies that in the generic case where the dark matter pair-annihilates and there are no symmetries or kinematic effects suppressing the natural pair-annihilation rate for a weak-scale weakly-interacting particle, the effect under investigation here is suppressed by 3 to 7 orders of magnitude compared to pair-annihilation. Notice also that the number of photons produced in each {\em annihilation event} is typically on the order of 10-100, as opposed to the single photon that scattering off of cosmic ray would produce {\em per event} (an exception to this is given by inelastic processes leading to neutral pions in the final state). Since the detectable quantity here is not the event-rate, but the actual gamma rays produced in the interactions, the suppression in the predicted gamma-ray flux is thus generically larger by a further 1 to 2 orders of magnitude.

If, however, effects such as co-annihilation or velocity dependence of the pair-annihilation are in place, the two processes could produce a comparable flux of gamma rays. In addition, if dark matter cannot pair-annihilate because of discrete symmetries or because of lack of annihilation partners, such as in asymmetric dark matter models \cite{Kaplan:2009ag}, then scattering off of cosmic ray electrons with a final-state photon emission might be one, if not the only way to detect a signal from dark matter with gamma-ray telescopes.

It is interesting to investigate two more dark matter scenarios that have received some attention recently: dipolar dark matter (see e.g. \cite{Banks:2010eh}), and luminous dark matter \cite{Feldstein:2010su}. The former arises as dark baryons in hidden-sector gauge mediation supersymmetry breaking (see e.g. \cite{Banks:2005hc}), or in technicolor models (see e.g. \cite{Nussinov:1985xr, Chivukula:1989qb, Chivukula:1992pn, Bagnasco:1993st}). While the relic density of this class of dark matter candidates is produced non-thermally, often a mechanism relates the dark baryon to the Standard-Model baryon density, thereby providing an explanation to the coincidence problem $\Omega_{\rm DM}\sim\Omega_{\rm B}$ \cite{Kaplan:1991ah}. Luminous dark matter, on the other hand, is an interesting proposal in light of recent direct detection results \cite{Feldstein:2010su}.

We estimate the dark matter scattering off of cosmic ray electrons for the case of dipolar dark matter assuming that the dark matter is a fermion $\psi$ of mass $m_{\rm DM}\sim 100$ GeV, with a dipolar interaction cross section of the type $${\cal L}_{\rm dipole}=\frac{e}{2\Lambda}\bar\psi\sigma^{\mu\nu}\psi F_{\mu\nu}.$$
We estimate the cross section off of cosmic-ray electrons with a typical energy scale around 1 GeV, and consider for simplicity the internal bremsstrahlung approximation of Eq.~(17) of Ref.~\cite{Gorchtein:2010xa} for the final state photon radiation, with an outgoing photon energy on the GeV range. Notice that this process is never resonant. We obtain the estimate $$\sigma_{e\chi\to e\chi\gamma}^{\rm dipole}\sim10^{-42}\ \left(\frac{1\ {\rm TeV}}{\Lambda}\right)^2{\rm cm}^2.$$ This entails that unless the dark sector scale $\Lambda$ is very low and the dark matter is very light the dipolar dark matter case is less promising than the WIMP case.

In the context of luminous dark matter \cite{Feldstein:2010su}, the dipolar operator connects two Majorana fermions, $\chi_g$ (the dark matter particle) and $\chi_e$, with $m_{\chi_g}\sim 1$ GeV and $m_{\chi_e}=m_{\chi_g}+\delta$ and $\delta\sim3.3$ keV chosen to provide an explanation to the observed DAMA/LIBRA spectrum. The relevant interaction term is $${\cal L}_{\rm lum}=\frac{i}{4\Lambda}\bar\chi_g\sigma^{\mu\nu}\chi_e F_{\mu\nu}\ +\ {\rm h.c.}$$
Here the picture is one where scattering of $\chi_g$ off of cosmic ray via the dipolar interaction operator produces meta-stable $\chi_e$ particles, subsequently decaying into $\chi_e\to\chi_g+\gamma$ with the photon energy in the final state $E_\gamma\sim\delta\sim3.3$ keV in the X-ray regime. 

We estimate the cross section for the process $e\chi_g\to e\chi_e$ for $E_e\sim1$ GeV to be on the order of $$\sigma^{\rm luminous}_{e\chi_g\to e\chi_e}\sim({\rm few})\times10^{-36}\left(\frac{1\ {\rm TeV}}{\Lambda}\right)^2\ {\rm cm}^2.$$ This cross section would produce a rate of 3.3 keV X-ray photons per unit dark matter target of $$q_{e\chi_g\to e\chi_g\gamma}\sim10^{-36}\left(\frac{1\ {\rm TeV}}{\Lambda}\right)^2\ \frac{1}{\rm s}.$$ Considering a Galaxy similar to the Milky Way, and taking a total mass of $M_{\rm MW}\sim10^{12}M_\odot$, this leads to an X-ray luminosity of about 
\begin{equation}
\label{eq:lum}
L_\gamma=\frac{M_{\rm MW}}{m_{\chi_g}}\ q_{e\chi_g\to e\chi_g\gamma}\sim10^{33}\ \frac{1}{\rm s},
\end{equation} 
which appears to be too small to be detectable if compared, for example, to the typical unresolved X-ray emission for a Milky Way-type galaxy of $$L_{\rm X-ray}\sim({\rm few})\times10^{39}\ \frac{\rm erg}{\rm s}\sim 10^{47}\ \frac{1}{\rm s}\ \ {\rm at}\ \ 1\ {\rm keV}.$$ Notice that the effective target number density is likely over-estimated in Eq.~(\ref{eq:lum}), as one should really only consider dark matter particles within the diffusion region where Galactic cosmic rays are confined. 

One can also compare our estimate in Eq.~(\ref{eq:lum}) with the luminosity that would be produced by sterile neutrinos with a lifetime $\tau\sim10^{31}$ s and a mass of a few keV, which lie at the verge of the sensitivity of current X-ray data (see e.g. \cite{Kusenko:2009up}). This case is relevant here since sterile neutrinos at that mass would produce exactly the same X-ray feature, a monochromatic line at an energy of a few keV. We estimate that the corresponding Milky Way X-ray luminosity would be on the order of $$L_{\nu_s}^{\rm lim}\sim10^{44}\ \frac{1}{\rm s}.$$
We thus infer that the luminosity we estimate for luminous dark matter excitation from cosmic-ray scattering, Eq.~(\ref{eq:lum}), is well below the current X-ray telescope sensitivity.

In what follows, we shall focus on the case of WIMPs, motivated by the estimates above, and by the fact that if resonances are present, those can produce interesting and potentially revealing spectral features, such as those we found in the case of scattering of dark matter off of AGN jet cosmic rays \cite{Gorchtein:2010xa}.

\section{Cosmic Ray - WIMP Dark Matter scattering}\label{core}
In this section, we use the analytical results of Ref.~\cite{Gorchtein:2010xa} on the neutralino-electron and neutralino-proton cross section with the radiation of a photon in the final state to calculate the diffuse gamma-ray galactic emission from dark matter scattering off of cosmic-ray electrons (sec.~\ref{sec:ele}) and off of cosmic-ray protons (sec.~\ref{sec:prot}).  
\subsection{Electrons}\label{ele}
\label{sec:ele}
We want to compute the spectrum of photons coming from the scattering of cosmic rays off of the dark matter in the Milky Way. We shall consider the electrons first. The photon flux is given by
\be 
\frac{dN}{dE_\gamma} = r_\odot \rho_\odot \bar J \frac{1}{M_\chi}\int d\Omega_\gamma \int dE_e \frac{d\phi}{dE_e} \frac{d^2\sigma}{d\Omega_\gamma dE_\gamma},
\ee
where 
\beqn
\bar J &=& \frac{2\pi}{\Delta \Omega} \int_{\Delta \Omega} d\theta \sin \theta J(\theta), \\ \nn
J(\theta) &=& \int_0^{2 r_\odot} ds \frac{1}{r_\odot \rho_\odot} \rho(r(s,\theta)) f(r(s,\theta)),
\eeqn
and $f(r) = e^{-r/r_0} / e^{-r_\odot /r_0}$ (see e.g. Ref.~\cite{Strong:2004td} and references therein) is included to take into account the fact that the cosmic ray flux is larger in the vicinity of the galactic center. In the expressions above, $\Omega_\gamma$ is the angle between the emitted photon and the incoming cosmic ray. To account for cosmic rays coming from every direction we integrate over the full solid angle $\Omega_\gamma$. On the other hand, $\Delta\Omega$ is the angle of observation from earth. We choose to focus  on the direction towards the Galactic Center (GC), where the dark matter density and the cosmic-ray density are highest; given the typical angular resolution, for the energies of interest, of the LAT \cite{Atwood:2009ez}, we choose an angular region of $\theta=1^\circ$ around the GC, which corresponds to $\Delta\Omega \sim 10^{-3}$.
As far as the dark matter distribution $\rho(r)$ is concerned, we use here the spherically-symmetric Einasto profile \cite{1989A&A...223...89E}
\be
\rho(r) = \rho_s \exp \left\{ -\frac{2}{\alpha} \left[\left(\frac{r}{r_s}\right)^\alpha -1\right] \right\}.
\ee
The cosmic ray electron flux is taken to be \cite{Ackermann:2010ij}
\be
\frac{d\phi}{dE_e} = k_e \left(\frac{E_e}{\rm GeV}\right)^{-3},
\ee
for $E_e$ between $\sim 0.1$ and $\sim 1000$ GeV. The numerical values for the various parameters we employ in the numerical evaluations of the Equations above are summarized in Table \ref{tab:param}.

\begin{table}[!b]
\begin{tabular}{|| l | c l ||}
\hline 
$r_\odot$ & {} & 8.33 kpc \\
$\rho_\odot$ &{}& 0.3 GeV cm$^{-3}$ \\
$r_0$ &{}& 4 kpc \\
$\alpha$ &{}& 0.11 \\
$r_s$ &{}& 35.24 kpc \\
$\rho_s$ &{}& 0.021 GeV cm$^{-3}$ \\
$k_e$ &{}& $10^{-2}$ GeV$^{-1}$ cm$^{-2}$ s$^{-1}$ sr$^{-1}$ \\
$k_p$ &{}& 1 GeV$^{-1}$ cm$^{-2}$ s$^{-1}$ sr$^{-1}$ \\
\hline 
\end{tabular}
\caption{The values in this table are from Ref.~\cite{Cirelli:2010xx}. \label{tab:param}}
\end{table}

The last ingredient is the differential cross section $\frac{d^2 \sigma}{d\Omega_\gamma dE_\gamma}$, in order to compute which, one has to specify a particle dark matter model. For the sake of illustration, we consider here the Minimal Supersymmetric Standard Model (MSSM), with the lightest neutralino as the dark matter candidate.
\begin{figure}[!t] 
\centering
\includegraphics[width=0.4\textwidth]{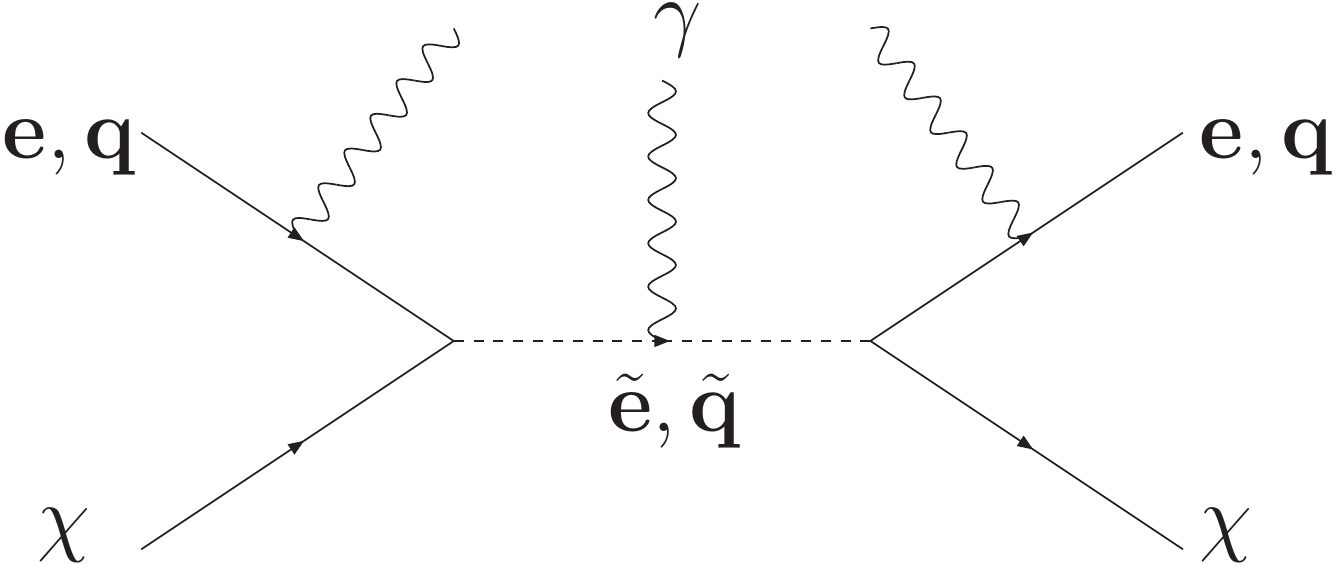}
\caption{\small\em Schematic diagram for electron (quark) - neutralino scattering proceeding through an $s$-channel.}
\label{fig:diagram}
\end{figure}

\subsubsection{Case study: Neutralino dark matter}
We recently carried out the complete calculation of the differential cross section in Ref.~\cite{Gorchtein:2010xa}, using the collinear approximation, {\em i.e.} taking the photon to be collinear with the final electron. We relax here, however, that assumption, while still only considering diagrams where the selectron is exchanged in the $s$-channel (see Fig.~\ref{fig:diagram}), since those are the only possibly resonant channels. Using the same notation as in \cite{Gorchtein:2010xa}, it is useful to remind the reader about the values of the coupling between electron (quark) - selectron (squark) - neutralino: $a_L = \sqrt{2} q g \tan\theta_W$ and $a_R = 1/2 a_L$, with $q$ the electric charge in units of $e$, $g$ the Standard Model $SU(2)$ gauge coupling and $\theta_W$ the Weinberg angle. $R$ and $L$ refer to the handedness of the electron or quark.

\begin{figure} 
\centering
\includegraphics[width=0.8\textwidth]{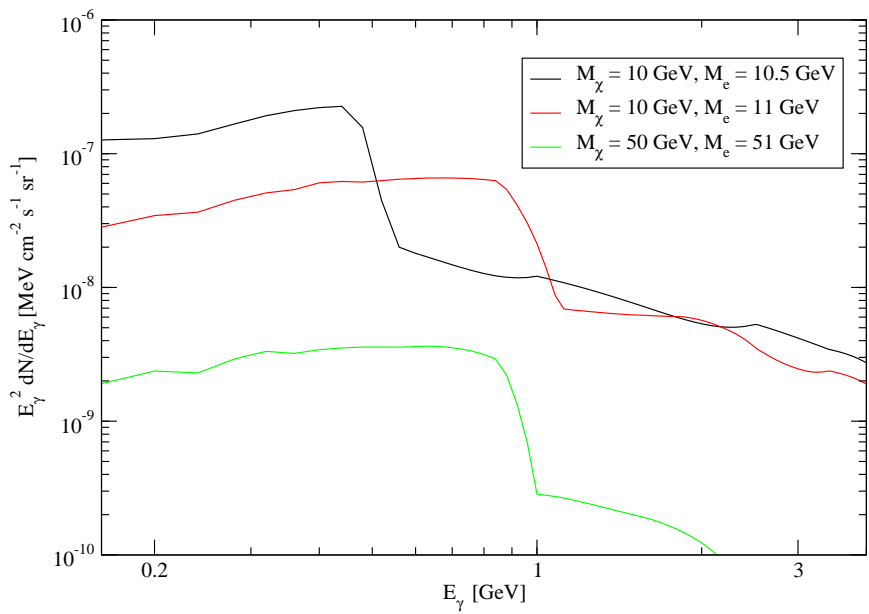}
\caption{\small\em The gamma-ray spectrum from cosmic-ray {\em electrons} scattering off of dark matter, for various dark matter particle masses $M_\chi$ and masses of the lightest electron-like scalar particle $M_e$.}
\label{fig:electrons}
\end{figure}

In Ref.~\cite{Gorchtein:2010xa}, the directions of the initial electron and the final photon were fixed by the direction of the AGN jet and its angle with respect to the line of sight. That made it appropriate to use the collinear approximation. Here the cosmic rays do not have a fixed incoming direction, thus we want to keep the full kinematics unconstrained, and we integrate over the full solid angle. In doing so, we encounter two distinct logarithmic enhancements: (i) when the photon is collinear with the final electron, as in \cite{Gorchtein:2010xa}, and (ii) when the photon is collinear with the initial electron. Notice that this integration is numerically non-trivial, as it runs over five independent variables, which can be taken to be the photon angle $\Omega_\gamma$, the direction of the final state electron, and the energy of the initial state electron. We adopt here a  Monte Carlo global adaptive integration strategy, as implemented in the {\tt Mathematica} package. 
 
We show our numerical results in Fig.~\ref{fig:electrons}, where we employ, for the sake of illustration, low lightest-neutralino masses (10 and 50 GeV) and small separations between the mass of the lightest selectron and the lightest neutralino (0.5 and 1 GeV). The mass difference sets the location of the kinematic condition that corresponds to the end of possible resonant effects. As a result, we find a sharp cutoff at 0.5 (black line) and at 1 GeV (red and green lines). Notice that the drop-off is of about one order of magnitude. Also, notice that the spectrum is significantly harder at energies below the kinematic drop-off, where it is slightly harder than $E_\gamma^{-2}$. In principle, this might make this particular process distinguishable from the diffuse galactic background. Also, from the standpoint of the angular distribution, this process is {\em not} isotropic, as the corresponding intensity correlates with the product of the cosmic-ray density times the dark matter density. Finally, we note that the level of intensity we predict for the signal for the models under consideration is only 3-4 orders of magnitude smaller than the isotropic diffuse background \cite{Abdo:2010nz}.

While a full-scale simulation of the detectability with Fermi LAT of the signal we calculated here is beyond the scope of the present analysis, we argue that the level of the gamma-ray emission predicted e.g. in the models illustrated in fig.~\ref{fig:electrons} might be potentially detectable. At a gamma-ray energy of around 1 GeV, in the same units as in fig.~\ref{fig:electrons}, the extragalactic diffuse emission is at the level of $\approx{\rm few}\times 10^{-4}$, and has a very regular power-law spectrum \cite{Abdo:2010nz}. Considering an angular region of interest of 1 sr, the LAT effective area of $\approx 10^4\ {\rm cm}^2$ at 1 GeV  \cite{Abdo:2010nz}, and an observing time of 3 years, $T\sim 10^8$ s, this corresponds to a total number of isotropic background photons collected at energies around 1 GeV on the order of $10^5$, and a total number of signal events (for e.g. the red line model of fig.\ref{fig:electrons}) of around $10^2$. This would correspond to a relatively low signal-to-noise (for example, $N_S/\sqrt{N_B}\sim$0.3), but in principle an excess corresponding to the sharp spectral feature we predict here could be detectable with Fermi LAT. A further handle would be to correlate the detected, albeit with low statistical significance, bump with its angular distribution, that should trace the product of the dark matter times galactic cosmic-ray density.

\subsection{Protons}\label{pro}
\label{sec:prot}
In the case of protons, the calculation is carried out at the parton level first, with a subsequent integral over the parton distribution functions  (PDFs) $f_i(x)$. We thus consider
\be
\frac{dN}{dE_\gamma} = r_\odot \rho_\odot \bar J \frac{1}{M_\chi}\int d\Omega_\gamma \int dE_p \int dx \sum_{i=u,d} f_i(x) \frac{d\phi}{dE_p} \frac{d^2\sigma}{d\Omega_\gamma dE_\gamma}(E_p, x),
\ee
where the flux of cosmic ray protons
\be
\frac{d\phi}{dE_p} = k_p \left(\frac{E_p}{\rm GeV}\right)^{-2.7}.
\ee
The calculation is analogous to the one done in the electron case. For numerical purposes, it is convenient to make the change of variables $E_q = x E_p$, with $E_q$ the energy of the incoming quark. Doing so, there is no dependence on $x$ in the differential cross section and the integrals over $x$ and $E_q$ can be done separately. 
\begin{figure} 
\centering
\includegraphics[width=0.8\textwidth]{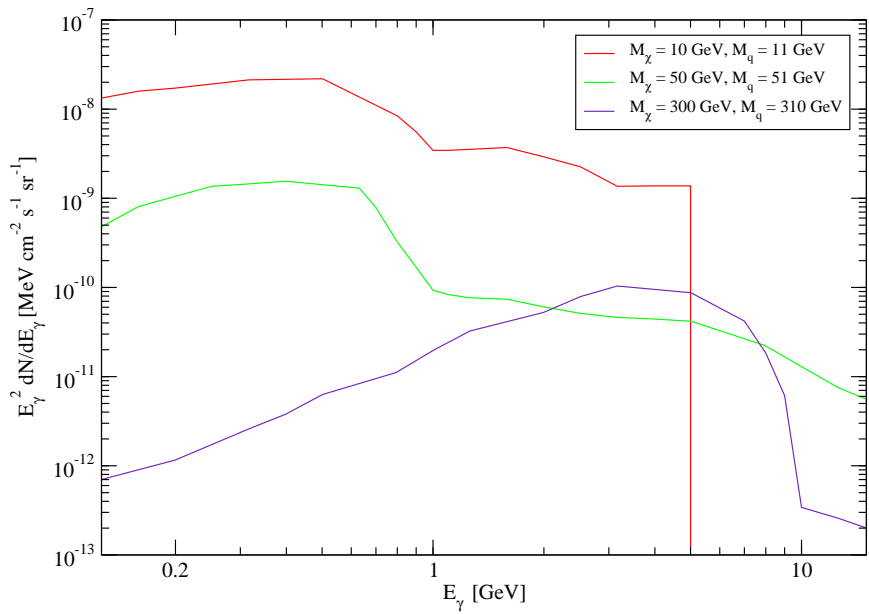}
\caption{\small\em The gamma-ray spectrum from cosmic-ray {\em protons} scattering off of dark matter, for various dark matter particle masses $M_\chi$ and masses of the lightest quark-like scalar particle $M_q$.}\label{fig:protons}
\end{figure}

As we see in Table \ref{tab:param}, the normalization for the cosmic-ay proton flux is two orders of magnitude higher than the one for electrons. Naively, one might thus hope to get a higher photon signal, but this is not actually the case. The coupling constants enter the cross section as $a^4_{L,R}$, and because of the fractional electric charge of the quarks, almost two orders of magnitude are lost there. This is of course only a simple model-dependent accident. The further integration over the PDFs, however, also suppresses the signal, because having partons at a given energy requires having much larger proton energies, and the spectrum of cosmic-ray protons falls very steeply with energy. As a result, we find that the photon flux coming from proton scattering ends up being comparable to the flux from electron scattering for the particular case of supersymmetric dark matter.

Using again for illustrative purposes the case of supersymmetry, we choose three sets of lightest neutralino-lightest squark mass pairs in Fig.~\ref{fig:protons}: $(M_\chi,M_q)=(10,11),\ (50,51),\ (300,310)$ GeV. Notice that, due to the integration over the parton distribution functions, the sharp drop-off we found for the cosmic-ray electron case is now smeared out over a broader range of photon energies, but is still noticeable. The overall normalization is still in a range which is, for some models, within 3-4 orders of magnitude of the isotropic gamma-ray background.

Notice that we find for the proton case a kinematical condition on the minimum quark energy which is completely analogous to the electron case \cite{Gorchtein:2010xa}
\be
E_q^{min} = \frac{E_\gamma}{1- E_\gamma/M_\chi (1-\cos \theta_\gamma)},
\ee
from which we see that the final photon energy cannot exceed $M_\chi /(1-\cos\theta_\gamma)$. That explains, for example, the sudden drop to zero at around 5 GeV for the line with $M_\chi=10$ GeV in Fig.~\ref{fig:protons}.

\section{Discussion and Conclusions}\label{sec:disc}
We studied the process of Galactic cosmic-ray electron and proton scattering off of dark matter with the emission of a photon in the final state. We argued that the gamma-ray flux from this process is, for WIMP models, suppressed compared to gamma rays from dark matter pair-annihilation. However, if the latter is in turn suppressed, or nonexistent as in asymmetric dark matter models, cosmic-ray-dark matter scattering can offer a unique opportunity for indirect dark matter searches with gamma rays. We find that, while for dipolar and luminous dark matter the intensity of the signal is well below detectability, for WIMP models resonances can enhance the signal to a detectable level in principle. 

We illustrated the expected photon spectrum for the particular case of supersymmetric dark matter, and for a few particle dark matter and selectron/squark masses. We found that the resonant structure entails a sharp spectral drop-off, at energies corresponding to the mass difference between the selectron/squark and the lightest neutralino. The spectrum is harder than $E^{-2}$ below the cutoff, and softer, and much more suppressed, at higher energies. In conclusion, while this process might be sub-dominant with respect to traditional searches for pair-annihilation signatures, it might still offer a unique chance of detecting indirectly dark matter with gamma-ray telescopes.

\begin{acknowledgments}
\noindent  The authors gratefully acknowledge conversations with Tom Banks, John March-Russell and Tomer Volansky. SP and LU are supported in part by an Outstanding Junior Investigator Award from the US Department of Energy and by Contract DE-FG02-04ER41268 (to SP). 
\end{acknowledgments}


\end{document}